# Mixing in superhydrophobic biomimetic rose petal passive microchannel

Prajakta Hiwase and Meenakshi V[1]

Department of Physics, Birla Institute of Technology and Science, Pilani – Hyderabad Campus, Hyderabad 500078, India

[1]E mail: meenakshi@hyderabad.bits-pilani.ac.in

**ABSTRACT**

Laminar flow prevails in microfluidic channels, and mixing is limited by the slow diffusive process at the interface between the mixing liquids. Whitesides and co-workers demonstrated a chaotic mixer by patterning the floor of the microchannel with obliquely oriented microridges. This creates an anisotropic resistance to flow, inducing exponential increase in stretching and folding of fluid elements leading to a dramatic increase in mixing. Rose petals that have micro papillae decorated with nanofolds, present a unique biomimetic, superhydrophobic, super adhesive surface. In the present work, a positive and negative replica of the rose petal surface was created using a simple, inexpensive soft lithographic technique and these replicas formed the floor of the microchannel. The floor of the channel naturally presents alternating slip and no slip boundary conditions for the fluid flow that in turn provides anisotropic drag reduction. We find that mixing index obtained for positive replica floor channel is distinctly higher than that of a negative replica floor channel. The variation in mixing index with Peclet number for the replicated channels are explained by characterizing their wetting behavior using contact angle and confocal microscopy. The enhancement in mixing due to the patterning of the floor is quantified by calculating the percentage change in mixing in the replicated channels from the corresponding simple Y channel values. The maximum percentage change is obtained for the positive replica floor channel (77.2 %) and compares well with chaotic mixer having obliquely oriented ridges on the floor (74.7 %).

## I. INTRODUCTION

The aim of micromixers is to mix two or more liquids completely in a short time over a short length of the microchannel, thereby enhancing the mixing efficiency. Rapid fluid mixing is important in terms of scientific interest and technological applications in various fields, such as drug delivery, protein folding, DNA sequencing, cell separation and detection, and environmental monitoring, to name a few [1]. The main challenge for mixing in microchannels is that it is diffusion-limited, occurring at the boundary between the two liquids, and requires very long channels (~ m) for complete mixing. Micro mixers are of two types - Active mixers and Passive mixers.[2] Active mixers rely on external sources of energy (like the electric field, magnetic field, thermal, acoustic, electrohydrodynamic, etc.)[3,4] and, in general, incorporate a mechanical transducer that stirs the fluids in the micromixer. Passive Mixers offer an attractive option in that they do not require an external power supply, and mixing, in general, is controlled by the geometry of the channel or by a surface pattern on channel walls.[3,4] The strategy that has been largely used to enhance mixing in microchannels is to create anisotropic resistance to flow that causes stretching and folding of fluid elements over the cross-section of the channel.

Pressure-driven flows (Poiseuille-like) in a simple microchannel, at low Reynolds number, lead to a parabolic velocity profile. Stroock et al.[5] altered this profile into a helical recirculation by patterning the floor of the Y microfluidic channel floor with a staggered herringbone-like structure as well as placing ridges inclined at an oblique angle on the floor: This presents an anisotropic resistance to flow between the ones traveling along the ridges and the ones traveling across the ridges. As a result, a transverse gradient develops, causing recirculation of fluid. Since this recirculation caused an exponential increase in the stretching and folding of fluid elements, the phenomenon is termed chaotic advection. Such chaotic flows reduce the mixing length dramatically from meters to a few centimeters.

In Stroock et al.'s [6] work, the floor of the microchannel is patterned in such a way that surface static charges on the floor of the channel alternate between negative and positive values in the same direction as an externally applied electric field. When an ionic fluid flows through this microchannel, it interacts with charge densities on the floor to create counter-circulating flows to enhance mixing. Here, the anisotropic 'resistance' to flow is provided by oppositely directed electric forces due to positive and negative charges.

Novel materials with special wettability have led to smart microfluidics that offers excellent friction control,[7-11] liquid reprography,[12,13] etc. Ou et al.[7] fabricated micron sized posts / micro ridges on the bottom wall of the channel by photolithography and they were made ultrahydrophobic using chemical treatment. Using confocal metrology system, they showed the trapping of air in the region between the microposts / microridges that lead to a shear free air –water interface; the pattern thus corresponds to an alternate stick (no slip) and slip boundary condition. The shear free interface reduces the flow resistance resulting in large slip velocities in microchannels by reducing the effective contact area between the liquid and solid surface. This idea was further extended by them [14] to create an anisotropic flow resistance by aligning the micro ridges (no slip) and hence the shear-free air-water interface (slip) supported between the micro ridge structures on an ultrahydrophobic surface at an oblique angle to the flow direction to enhance mixing. Optical images showed helical secondary flow patterns that stretched and folded the fluid elements and reduced the mixing length as compared to a smooth Y channel. The design of these hydrophobic surfaces in a Y channel was optimized by numerical simulation. Changing the width of microposts or the separation between them did not affect mixing. Vagner et al. [15] studied the mechanism behind helicoidal flow formation in a microchannel with super hydrophobic walls and found that they originated from variations in transverse pressure gradients. Topological features incorporated in super hydrophobic surfaces provide significant drag reduction in micro channels as investigated by Martell et al.[16] using numerical simulation. They found that slip velocity primarily depends on the micro feature spacing for configurations with Reynolds number 180 – 590.

The dynamics of wetting relevant to microfluidics and nanofluidics is discussed in detail by David Bonn et al.[17] In general, the wetting behavior of a droplet on a superhydrophobic surface can be described in terms of Wenzel model and Cassie-Baxter model.[18,19] In the Wenzel model,[20] the liquid completely wets the rough surface. In the Cassie-Baxter model,[21] there is a liquid/air/solid composite interface between the liquid and solid. Significant scientific interest and a variety of potential applications make superhydrophobic surfaces interesting.[22] Smart surfaces can switch wettability in response to a variety of environmental stimulus.[23-26] Micron and sub-micron scale hierarchical roughness in these surfaces influences the hydrophobicity. [27-31] Contact angle (CA) quantifies the affinity between the liquid and solid surface while contact angle hysteresis (CAH) quantifies the mobility of the liquid on the solid surface.[32] The mobility of a liquid drop on a solid surface can alternatively be characterized by the tilting angle of the solid flat surface at which the drop starts to slide off. Superhydrophobic surfaces have CA > 150 and a low CAH (< 5º or 10º) / low sliding angle.

Dey et al.[33] studied the hydrodynamic characteristics of microchannel with a lotus leaf replica wall; lotus leaf is known to be superhydrophobic as a consequence of the hierarchical structures on its surface and is less adhesive. While stick – slip dominated flow over the superhydrophobic surface is expected, they observed two distinct flow regimes even at low Reynolds number that is governed by the flow rate. It was observed that mixing is enhanced in micro channels with biomimetic lotus leaf replica wall.[34] The stick - slip flow over a super hydrophobic surface does not exist throughout the low Reynolds number regime. Micro particle image velocimetry experiments (as visualized from velocity vector maps) revealed that when the flow rate exceeds a critical value, an enhanced mixing is observed due to the change in flow characteristics: the interfacial condition changes from Cassie-Baxter (stick – slip) to Wenzel (no slip) state. The results indicated that the biomimetic rough surface alters the laminar flow to an erratic flow at a critical flow rate that also depends upon the confinement ratio (ratio of protrusion length of the lotus replica to the hydraulic diameter of the channel). Increasing confinement ratio enabled a faster transition to the erratic flow regime.

Our work aims to understand the role of boundary condition induced flow anisotropy in enhancing mixing in passive micromixers through experiments at low Reynolds number (5 – 75). To this end, we use positive and negative replica of rose petal super hydrophobic (highly adhesive in contrast to a lotus leaf) surface for the first time as the floor of microfluidic Y channel mixer. For comparison, a simple Y channel with a flat floor is also fabricated and 3D numerical simulations are done for the simple Y channel. There is limited literature available for passive micromixers designed with superhydrophobic surfaces where experimental mixing efficiency is quantified, although superhydrophobic surfaces are known to cause drag reduction. [7,14,15,33,34] Most of the fabrication of micromixers with superhydrophobic surfaces, with an exception of few [33,34,35] involve a multistep photolithographic process that limit the fabrication in the absence of these sophisticated expensive facilities. Our fabrication involves a simple, two-step inexpensive process using soft lithography[36] with the red rose petal surface serving as the master and the Y channel is fabricated with a laser engraving machine.

## II. EXPERIMENTAL AND SIMULATION METHODOLOGY

### A. Microchannel fabrication

In order to study the mixing characteristics with varying boundary conditions (flat surface, negative and positive replica of rose petal) on the floor of the microchannel, three Y micro channels are fabricated. The advantage of the present method is that it is simple and does not require any sophisticated instrument for its fabrication. The geometry of a simple Y microchannel is shown in Fig. 1. The length, width, and height of the microchannel are 25 mm, 1 mm, and 0.5 mm, respectively. The microchannel was designed (negative to the actual channel) using AUTOCAD software with two inlets and one outlet and fabricated on PMMA (poly methyl methacrylate) using a 100-watt $CO_2$ laser engraver that serves as a master. Soft lithography[36,37] with Poly di methyl siloxane (PDMS), a thermo curable polymer from Sylgard-184 (Dow Corning) that comes as a two component kit – (i) silicone base and (ii) curing agent, is used to fabricate the microchannel. A mixture with 10:1 ratio of pre-polymer to curing agent is used for all replications. The degassed PDMS mixture is poured on the fresh rose petal and cured at 35⁰C for 36 hours. The solidified elastomer is then peeled off from the petal surface in order to get the PDMS mold. The Y patterned PDMS mold is placed with its patterned surface down on a flat PDMS of similar dimension that serves as the floor (smooth base) of the simple Y channel and mechanically clamped to create a leak proof simple Y microfluidic channel. Two more channels are fabricated with identical dimensions as shown in Fig. 1, but the negative and positive replica (as detailed in the next section) of rose petal is used as the floor of the channel.

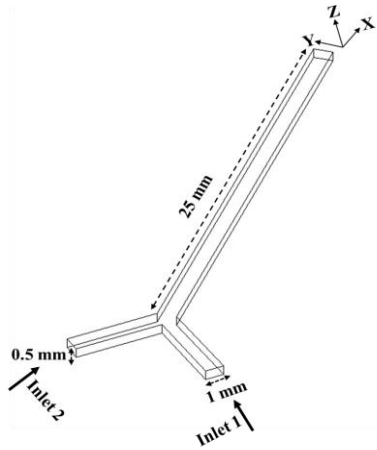

**FIG. 1.** Schematic of a simple Y channel

A two-step replica molding process was followed with steps outlined in Fig. 2 to replicate the pattern in the rose petals on PDMS and the procedure is similar to the one adopted by Ghosh et al.[38] The first step of replica molding is to obtain the negative replica of the rose petal, which is then used as master to obtain the positive replica. Fresh rose petals were stuck on a petri dish using double-sided sticky tape. In order to maintain the same thickness of the PDMS in all the channels, a predetermined amount (0.1 g) of degassed Sylgard-184 prepolymer solution with curing agent is poured over the rose petal. The PDMS is then cross-linked by placing it in a hot air oven at 35ºC for 36 hours. The low curing temperature ensures that rose petals are not damaged in the process. The cross-linked PDMS is peeled off from the rose petal after curing to obtain the negative replica of the rose petal. For the second stage of replica molding, the negative replica acts as the master template. The negative replica is kept in a UVO chamber (Novascen, USA) and subject to UVO exposure to inhibit cohesive adherence between the stamp (negative replica) and the PDMS layer to be patterned. The surface is now exposed to oxygen radicals, which in turn generates a superficial oxide layer that prevents the PDMS negative replica from sticking to PDMS in the next replication process. Subsequently, PDMS mixed with the curing agent is poured on the negative replica and annealing is done for three hours at 100º C to cross-link the PDMS layer completely [38] and the positive replica is then peeled off. The positive (negative) PDMS replica served as the floor of the channel for the positive (negative) patterned floor channel; The PDMS Y channel obtained from soft lithography was mechanically clamped on to the positive (negative) replica to form a leak proof microfluidic channel with a patterned floor.

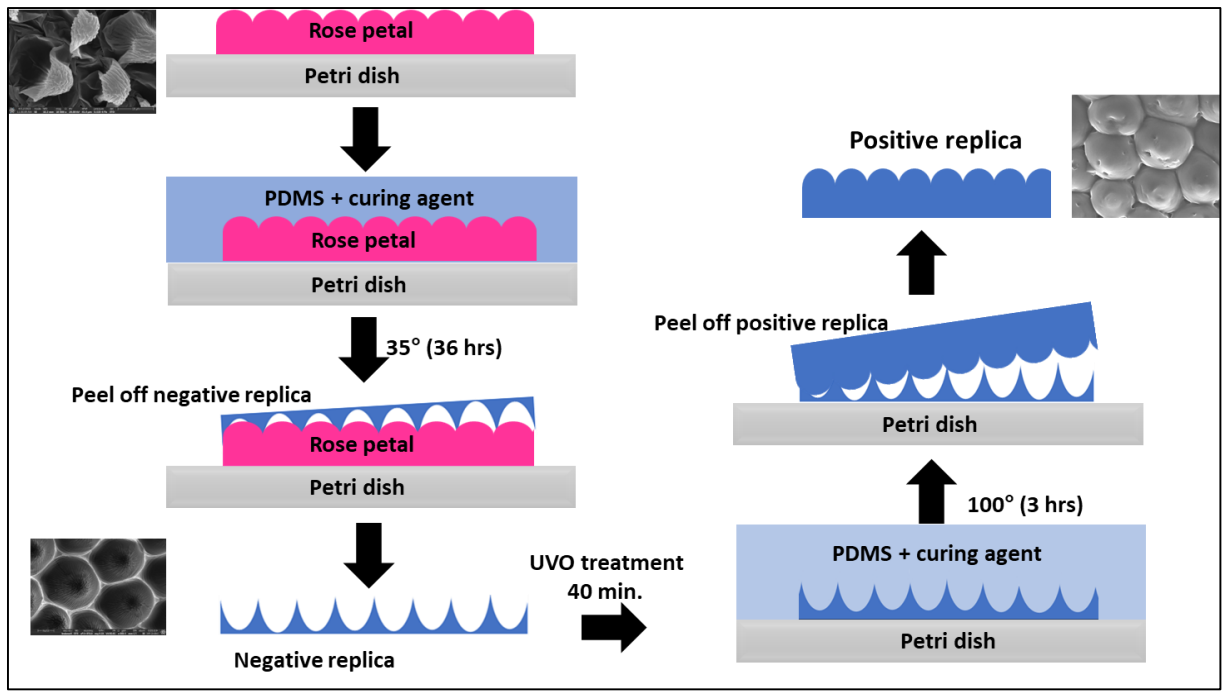

**FIG. 2.** Steps in replication process to obtain the positive and negative replica. SEM images of the rose petal, positive and negative replica are also shown alongside the schematic.

### B. Computational Modeling

Numerical simulations in 3D were performed using COMSOL Multiphysics software (version 6.0 a) with microfluidics module for a simple Y mixer. The governing equations used to model the three-dimensional Newtonian incompressible flow in the channel are continuity equation (Eqn 3), Navier – Stokes equation (Eqn 4), and convection - diffusion equation (Eqn 5).

$$\frac{\partial \rho}{\partial t} + \nabla \cdot (\rho \mathbf{v}) = 0 \qquad (3)$$

For incompressible flow, $\nabla \cdot \mathbf{v} = 0$

$$\rho \left[\frac{\partial \mathbf{v}}{\partial t} + (\mathbf{v} \cdot \nabla)\mathbf{v}\right] = -\nabla P + \mu \nabla^2 \mathbf{v} \qquad (4)$$

$$\frac{\partial C_i}{\partial t} + \mathbf{v} \cdot \nabla C_i = D_i \nabla^2 C_i \tag{5}$$

where, ρ, μ are the density and dynamic viscosity of the fluid; $C_i$ and $D_i$ are the concentration and the molecular diffusion coefficient of the species; t and P are the time and pressure, respectively, and **v** is the velocity field. Finite element method is used to solve the above governing differential equations by recursively computing them with the given (user) boundary conditions in the various discretised fluid elements. Mesh Convergence showed a ± 4% error in mixing index values between the various meshes (Number of Elements: Extremely fine (4656727), Extra fine (4192645), Finer (924258), Fine (263111), Normal (100910), Coarser (21369)). Simulations are performed with physics controlled fine mesh (263111 elements). Fine mesh was chosen after performing multiple simulations to optimize the mesh elements. The channel dimensions used for the simulation and experiments are identical. Water (concentration 0 mol/m³) is sent through inlet 1 and Rhodamine B dye (concentration 1 mol/m³) through inlet 2 (Fig. 1) with a diffusion coefficient (D) of $4.5 \times 10^{-10}$ m²/s. Four simulations are done for a simple Y channel with mean inflow velocity of 0.01 m/s, 0.05 m/s, 0.1m/s and 0.15 m/s. The density (ρ) and viscosity (μ) are taken as 1000 kg/m³ (water) and 0.001 Pa. s, respectively. The boundary conditions used for a simple Y channel are no-slip at the channel walls and pressure at the outlet was maintained at zero static pressure. Reynolds number (Re) was calculated using Re = ρvl/μ; l is the characteristic length of the channel. In this study, we considered the depth to be the characteristic length. Experiments and Simulation were done for Reynolds number of 5, 25, 50, 75 (calculated using depth) by varying the inlet velocity of the flow in the channel. Peclet number (Pe) defined as the ratio of convection to diffusion rate is calculated using Pe = vl/D. The standard deviation of concentration values across the channel width is evaluated to estimate the mixing index for the results from numerical simulation.

$$\text{Mixing index} = 1 - \sqrt{\frac{1}{N} \frac{\sum_{i=1}^{N}(C_i - \overline{C})^2}{\overline{C}}} \tag{6}$$

where N is the total number of data points taken across the cross-section, $C_i$ is the concentration at every individual data point, $\overline{C}$ is the average concentration value over the complete data points. COMSOL's 'cut point' dataset feature is used for extracting the concentration data across the cross-section of the channel at 0.2 mm before the end of the channel. The concentration values are exported as .txt file and imported to an in-house MALAB code to calculate mixing index at the end of the channel.

Model validation was done by performing simulation for an identical Y channel (3D) used by Vagner et al. 2019 [15]. Figure 3 gives a clear indication that the data from our simulation (circular symbols) is consistent with the data of Vagner et al. 2019 (square symbols).[15] The mixing index computed at the end of the channel from our simulation is 90.9%, which is in agreement with the reported value of 89.5%.

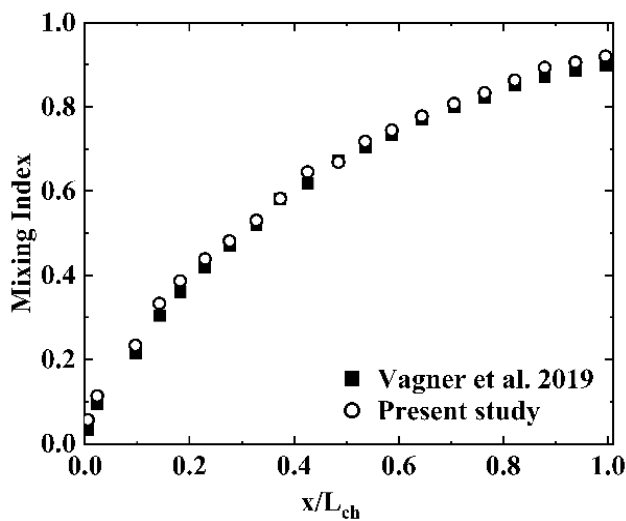

**FIG. 3.** Variation in mixing index across the cross-section at various positions (x/$L_{ch}$) along the length of the channel for a simple Y channel, where $L_{ch}$ is the channel length. Open circular Symbols are the data points from the present simulation, and solid rectangular symbols are data from Vagner et al. 2019 [Fig. 17(b)]

### C. Characterization Techniques

In order to understand the surface topology and confirm the transfer of features from rose petal to the positive and negative PDMS replica, Field Emission Scanning Electron Microscopy (FESEM, Oxford X-max LEICA EM ACE200) was performed on fresh rose petal, negative and positive PDMS replica. The samples were coated with a thin conducting layer of gold and scanned in the high vacuum mode at 20 kV.

For the quantification of wetting state of rose petal, its negative and positive replicas, contact angle measurements were performed using a mobile phone camera (Redmi mobile phone using 8X magnification).[39] The samples are placed on a flat surface. Deionized water droplets of constant volume (5 μL) were dispensed using a micropipette at multiple locations and the average contact angle is determined using Image J software.

Confocal Microscopic studies are done to understand the wetting states of positive and negative PDMS replica as they will be used in the floor of the microfluidic channel for mixing studies. An inverted laser scanning confocal microscope (Leica DMi 8) with a 10X objective was used for this study to scan the xy plane. To determine the exact nature of the solid / liquid interfaces, a 10 μL droplet of Rhodamine B (0.667 g/l) dyed water was dispensed on the surface. The interface distinction was understood based on the color of the surface: regions where the Rhodamine dye can enter looks red, whereas the darker region shows the absence of Rhodamine dye.

Atomic force microscopic measurements (AFM) were done to confirm the height of the micro and nano structures on the positive and negative replica of the rose petal surface.

**D. Mixing Index experiments**

Experiments were performed on three different Y microfluidic channels (simple Y channel, Negative replica floor (bottom wall) Y Channel and Positive replica floor Y channel), all having the same dimensions as outlined in microchannel fabrication. The setup used for flow visualisation consists of a dual channel programmable syringe pump (New Era 4000X, USA), a fluorescent microscope (Olympus BX43) equipped with a CCD camera (SCMOS pco edge), and a monitor with software to record the captured frames. Two syringes are connected to the inlets of the micro channel (Fig. 1) with the help of tubing and needles. The liquids used at the inlets are water (top inlet) and Rhodamine-B dye (bottom inlet) with 1 millimolar concentration. The syringes are manually fixed to the programmable syringe pump and flow rates are suitably set to achieve the desired Reynolds number (Re) in order to enable a direct comparison with the simulation results. All experiments are done using a 60 ml syringe and flow rate varied to achieve the desired velocity / Re. Images were captured at 4X magnification for a region of 3.4 mm before the outlet of the channel to calculate mixing index. The experiments were performed for four different velocities i.e., 0.01 m/s (117 µl per minute), 0.05 m/s (588 µl per minute), 0.1 m/s (1172 µl per minute) and 0.15 m/s (1764 µl per minute). Each experiment, was repeated thrice to ensure repeatability and for each experiment, average of ten images are taken for calculating mixing index. Mixing Index was calculated using the pixel intensities from the images at 0.2 mm before the end of the channel using the following equation[40,41]

$$\text{Mixing Index (MI)} = 1 - \frac{\sqrt{\frac{1}{N}\sum_{i=1}^{N}(I_i - \langle I \rangle)^2}}{\sqrt{\frac{1}{N}\sum_{i=1}^{N}(I_{0i} - \langle I \rangle)^2}} \qquad (7)$$

where, N is the total number of pixels across the cross section, $I_i$, $\langle I \rangle$ and $I_{oi}$ is the intensity at every pixel, average pixel value, intensity of unmixed fluid (at the pixel i). All the quantities listed in the Eqn. 7 were experimentally determined.[42] The above equation computes the ratio of standard deviation of pixel intensities across a cross section to the standard deviation of pixel intensities in the unmixed case and a inhouse matlab code was used to calculate the mixing index.

## III. RESULTS AND DISCUSSION

Figure 4(a), 4(b) and 4(c) shows the SEM images of rose petal, negative replica and positive replica, respectively. The SEM image reveals the hierarchical structure of the rose petal and confirms the presence of micropapillae that are covered with the nanofolds (see inset figure 4(a)) on their surface, which are densely packed on the top. The typical honeycomb structure of the negative replica is confirmed in Figure 4b. The negative replica, which is the reverse structure of the rose petal, has a microscopic well in the place of micro papillae and has nanofolds inverted on the bottom of the well. Figure 4(c) is the positive replica microstructure, which is obtained after the double replication process (refer to Figure 2). The positive replica features are expected to be identical to that of rose petal microstructures. Image J software is used for the measurement of dimension in SEM images. Table I compares the average dimensions of the micro and nanostructures of papillae from this work with the existing literature.

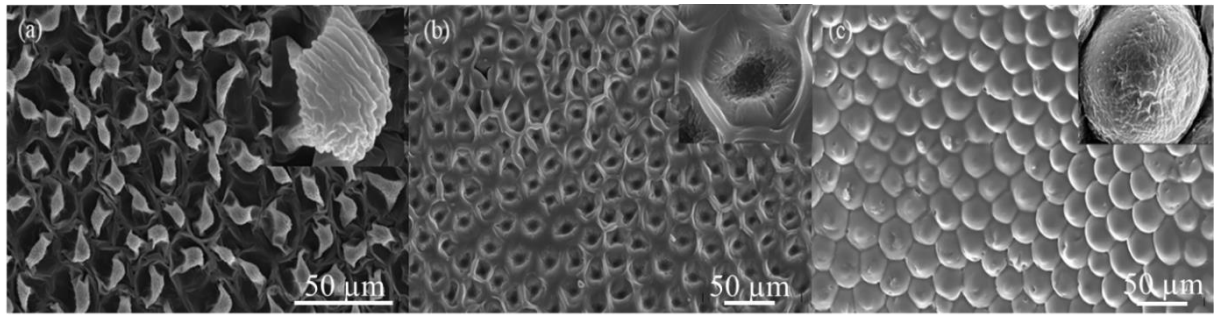

**FIG. 4.** FE SEM images of (a) Rose petal (Magnification: 2000X) (b) Negative replica (Magnification: 1000X) (c) Positive replica (Magnification : 1000X). The insets show the zoomed microstructure over the one papilla.

TABLE I. Micro and nanostructure dimensions obtained using SEM characterization

| Dimension | Positive replica (Literature) | Positive replica (Present study) | Negative replica (Present study) |
|---|---|---|---|
| Micropapillae diameter | 18 - 30 μm[38,43,44] | 29.60 ± 2.70 μm | 22.2 ± 2.7 μm |
| Distance between nanopillars | ~1 μm [45] | 0.32 ± 0.06 μm | - |
| Distance between micropapillae (peak to peak) | ~30 μm [38] | 31 ± 3 μm | 4.5 ± 0.7 μm |
| Nanopillars diameter | 0.3 μm ~ 0.5 μm[46] 0.73 μm [47] | 0.45 ± 0.16 μm | - |

Results of the micro structure from SEM agrees well with other reported literature. [1,47,48] The contact angle measurements were repeated thrice and average contact angle obtained on the rose petal and its negative and positive replica for 5 µL drop volume were found to be 146.0 ± 0.2°, 144 ± 1°, and 147 ± 1°, respectively (Table II). This study confirms the hydrophobic nature of the negative and positive replicas. The contact angle was calculated using Image J software. The contact angle measurement studies show that the contact angle for negative replica is marginally lower than that for positive replica.

TABLE II. Contact angle measurements

| Sample | *Present study* | *(Ghosh et al. 2019)*[38] | *(S. Dai et al.2019)*[49] |
|---|---|---|---|
| Fresh rose petal | 146.0 ± 0.2° | 140.0 ± 5.0° | 152° |
| Positive replica | 147.0 ± 1.0° | 138.5 ± 2.6° | 153° |
| Negative replica | 144.0 ± 1.0° | 136.8 ± 3.0° | 148° |

The wettability of the replicated surfaces was studied using confocal microscopy. Figure 5 shows the images captured using a confocal microscope for (a) negative replica and (b) positive replica. In the negative replica, high intensity of Rhodamine dye at the hexagonal edges of the micropapillae shows that the Rhodamine could enter only the edges. The micro well region is dark, indicating the absence of dye within the micro-wells. This indicates that negative replica has air-filled microwells, corresponding to the Cassie Baxter wetting state. [38] However, in the positive replica, the slim region between micropapillae are dark as Rhodamine has not entered but there is a variable contrast in red color on top of micropapillae. This is indicative that wetting happens on the top of micropapillae and has reduced wetting in the regions below the top.

Atomic force microscopy was performed to confirm the micro and nano structures on the negative and positive replicas and the depth of the micro well in negative replica was found to be ~ 6 µm and the height of the micropapillae in positive replica was also ~ 6 µm.

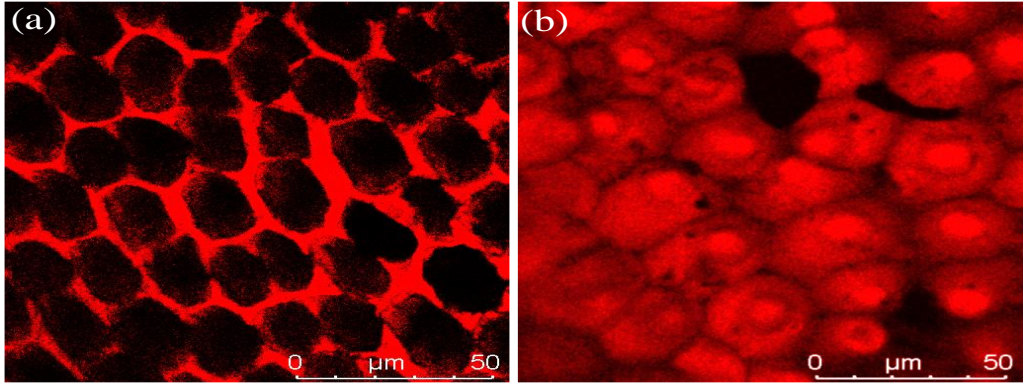

**FIG.5.** Confocal images of (a) Negative replica (b) Positive replica, showing the wetting state.

Figure 6 shows a plot of experimentally determined mixing index (MI) for the three different microfluidic channels at 0.2 mm before the outlet of the channel as a function of Peclet number. The average of the mixing index obtained from three different measurements performed for each type of microfluidic channel is shown and standard deviation of the three measurements is used for estimating the error shown in the graph. Data from simulation (filled star data in Fig. 6) for a simple Y channel is also shown for comparison and it agrees well with the experimental data. The channel with the positive replica floor has the highest mixing index and the jump in mixing index occurs at an inlet velocity of 0.05 m/s. Both the positive and negative replica floor channel indicate a saturation of the mixing index beyond an inlet velocity of 0.10 m/s. In the case of negative replica, for the lower Pe, the mixing index values are coincident with that of a simple Y channel. The transition / jump to a higher mixing index occurs at a lower flow rate (0.01 m/s velocity) for the positive replica as compared to the negative replica floor channel (0.05 m/s velocity).

To explain the difference in mixing index, we need to understand the wetting behavior in a rose petal which is discussed below. Jin et al.[49] investigated the dynamic wetting characteristics of rose petal using optical microscopy and proposed a geometric model with three co-existing wetting states: (i) air cushion covering multiple micropapillae which exhibits 'Cassie state with air cushion'; this is different from Cassie impregnating state where the liquid wets the large scale micro grooves and air exists in smaller ones, (ii) classic Cassie state and (iii) Wenzel state. The reason for state (i) is attributed to the fact that when surface defects cause wetting of micro papillae, the displaced air gathers along with the already accumulated air among other micro papillae, resulting in the growth of the air cushion that cover multiple micropapillae, and are sealed when they are away from the interface. The sealed air cushion results in the strong adhesive force.

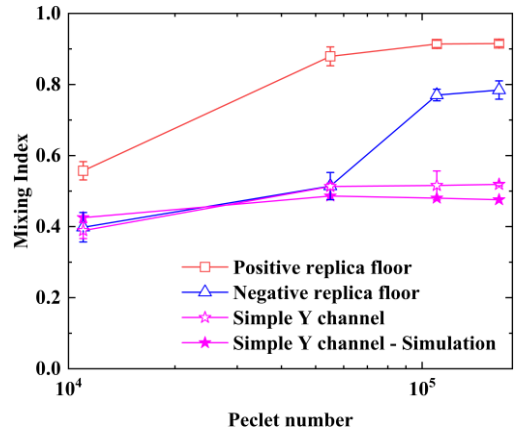

**FIG. 6** Plot of Mixing Index as a function of Peclet number at 0.2 mm before the outlet of the channel for the three different microchannels. Simulation data is also shown for simple Y channel for comparison.

**TABLE III.** Percentage change in Mixing index. (*indicates that the diffusion coefficient used is one order higher)

| Pe | Mixing index of simple Y channel | Percentage change in Mixing index (%) Positive replica | Negative replica |
|---|---|---|---|
| 11000 | 0.39 | 43.3 | 2.4 |
| 55000 | 0.51 | 71.4 | 0.2 |
| 110000 | 0.52 | 77.2 | 49.3 |
| 165000 | 0.52 | 76.4 | 51.1 |

| Pe | | Oblique Ridges [5] | Staggered Herringbone mixer [5] |
|---|---|---|---|
| 200000 | 0.51 | 74.7 | 94.4 |

| Pe | | Microridges Fig 3 of Ou et al. [14] |
|---|---|---|
| 435 | 0.29 | 220.7* |
| 2175 | | 107.0* |
| 4350 | | 49.6* |

Teisala et al.[50] argue that the geometry of fine structure on a hierarchically rough surface determines the adhesion of a surface. While in the case of lotus leaf surface, spiky or tubular like fine structure reduces the liquid-solid contact area and adhesion, in contrast, a rounded fine structure increases the liquid-solid contact area that can also exceed that of a smooth surface. A round and grooved fine structure as seen in rose petal further increases the liquid-solid contact area, anchoring the water droplets strongly to the surface resulting in high adhesion. In this case, water was seen to only wet the top regions of micro-papillae and air is trapped in the region between micro-papillae and also in the nanofolds.

The negative replica floor channel where wetting happens only at the edges of hexagons (Fig. 5a) has a slim liquid-solid contact area but large air filled hexagonal pockets as evidenced from confocal microscopy and there is no anisotropic resistance to the flow. In the absence of a recirculating flow contribution arising from the channel floor, the remaining region only contributes to diffusion and this is identical to that of a simple Y channel. When the inlet flow rate is increased corresponding to a velocity of 0.05 m/s (Pe 55000), a sudden jump in the mixing index is observed.

For a positive replica floor channel, even at the lowest Pe (velocity 0.01 m/s), the mixing index value is significantly higher than that of the negative replica floor channel, and a jump in mixing index happens at a velocity of 0.01 m/s (Pe 11000) indicating a transition in the flow behavior. As seen in the confocal microscopic image (Fig. 5(b)) and as suggested by Teisala et al.,[50] wetting happens on the top of micropapillae predominantly and there are air pockets within the nano pillars on the surface of the papillae. It is known that shear-free air water interfaces have large slip velocities.[14,7] The alternating slip and no slip (stick) boundary condition on the micro papillae surface may create a phase lag between the two flows that results in transverse flow components that causes a jump in mixing. Another plausible reason could be the roughness on the surface of micro papillae leads to more shearing of the liquid, that enhances mixing.

One basic difference between the positive and negative replica floor channel is that the micropapillae height projects ~ 6 μm (from AFM) above the bottom wall of the channel making the effective depth of the channel to be 494 μm whereas the microwell in negative replica is ~ 6 μm deep from the surface making the effective channel depth 506 μm. The difference in effective depth partially accounts for the difference in mixing index values between the positive and negative replica floor channels observed at all flow rates (Pe).

To account for the sudden jump in the mixing index observed for the replicated channels, the area of the shear free air - water interface present in the microwells of the negative

**14**

replica floor channels are much larger than the shear free air-water region in the positive replica floor channels. More energy is required to displace the air from the microwells of the negative replica floor channels and hence the jump in the mixing index occurs at a higher flow rate (Pe 55000) as compared to the positive replica floor channel.

The mixing index values for both positive replica floor channel and negative replica floor channel tend to a saturation value at Pe 55000 and Pe 110000, respectively. This would correspond to the depinning of the three phase contact line and the Wenzel wetting state setting in.

The enhancement in mixing in the positive and negative replica floor channels has been quantified by calculating the percentage change in mixing $\left(\frac{MI_{replica} - MI_{simple\,Y}}{MI_{simple\,Y}}\right)$ using the corresponding simple Y channel mixing index values. We have taken data from Stroock et al.'s [5] paper for channel with obliquely oriented ridges in the floor and staggered herringbone mixer and Ou et al.'s [14] data for the channel with microridges to calculate the percentage change in mixing from their respective simple Y channel mixing index value. Table III lists the percentage change in mixing index obtained from the present experiment as well as the work by Stroock et al. and Ou et al. The maximum percentage change obtained from the present experiment is seen for the positive replica floor channel (77.2 %) for a Pe of $1.1 \times 10^5$ and that compares well with channels with obliquely oriented ridges[5] on the floor (74.7 %) with Pe $2 \times 10^5$. The results suggest that further optimization of the channel geometry by simulation and experiments will lead to further enhancement in mixing.

## IV. CONCLUSIONS

This study for the first time exploits replicated rose petal surface for enhanced mixing. We have used an inexpensive method to replicate the rose petal surface. We compared the results of the positive and negative replica floor channel with other reported literature having micro ridges and herringbone on the floor of the microchannel. We found that positive and negative replica floor channel have markedly different mixing indices and we explain this using the alternating stick / slip boundary condition that is more effective in positive replica. Future studies will be directed towards further modifications to this geometry to create chaotic advection.

## ACKNOWLEDGMENTS

We thank Prof. Aravinda N Raghavan for continuous interaction and helpful comments. We thank the Central Analytical Lab for confocal microscopy.

## AUTHOR DECLARATIONS
**Conflict of Interest**

The authors have no conflicts to disclose.

## DATA AVAILABILITY

The data that supports the findings in this study are available from the corresponding author upon request.